\documentclass[aps,prc,reprint,showpacs,groupedaddress,twocolumn]{revtex4-1}

\usepackage{graphicx,color} 
\usepackage{dcolumn}  
\usepackage{bm}       
\usepackage{amsmath}

\begin{document}

\newcommand {\nc} {\newcommand}

\newcommand{\vv}[1]{{$\bf {#1}$}}
\newcommand{\ul}[1]{\underline{#1}}
\def\bsigma{\mbox{\boldmath$\sigma$}}

\nc {\IR} [1]{\textcolor{red}{#1}}
\nc {\IB} [1]{\textcolor{blue}{#1}}
\nc {\IP} [1]{\textcolor{magenta}{#1}}

\title{Separable Representation of Proton-Nucleus Optical Potentials} 

\author{L.~Hlophe$^{(a)}$}
\email{lh421709@ohio.edu}
\author{V.~Eremenko$^{(a,e)}$}
\author{Ch.~Elster$^{(a)}$}
\email{elster@ohio.edu}
\author{F.M.~Nunes$^{(b)}$, G.~Arbanas$^{(c)}$, 
J.E.~Escher$^{(d)}$, I.J.~Thompson$^{(d)}$
}

\affiliation{(a)Institute of Nuclear and Particle Physics,  and
Department of Physics and Astronomy,  Ohio University, Athens, OH 45701 \\
(b) National Superconducting Cyclotron Laboratory and Department of Physics and Astronomy, Michigan State University, East Lansing, MI 48824, USA \\
(c) Reactor and Nuclear Systems Division, Oak Ridge National Laboratory, Oak Ridge, TN 37831, USA \\
(d) Lawrence Livermore National Laboratory L-414, Livermore, CA 94551, USA
(e) D.V. Skobeltsyn Institute of Nuclear Physics, M.V. Lomonosov
Moscow State University, Moscow, 119991, Russia
}

\collaboration{The TORUS Collaboration }
\noaffiliation

\date{\today}

\begin{abstract}
Recently, a new approach for solving the three-body problem for (d,p) reactions in
the Coulomb-distorted basis in momentum space was proposed. Important input
quantities for such calculations are the scattering matrix elements for proton-
and neutron-nucleus scattering. We present a generalization of
the Ernst-Shakin-Thaler scheme in which a momentum space 
separable representation of proton-nucleus scattering matrix elements 
can be calculated in the Coulomb basis.
The viability of this method is demonstrated by comparing S-matrix elements obtained
for p$+^{48}$Ca and p$+^{208}$Pb for a phenomenological optical potential with
corresponding coordinate space calculations.
\end{abstract}


\maketitle

\section{Introduction}
\label{intro}
\vspace{-3mm}

Deuteron induced nuclear reactions are attractive from an experimental as
well as theoretical point of view for probing the structure of exotic nuclei and as 
an indirect tool in astrophysics (see e.g.~\cite{Kozub:2012ka}). 
From a theoretical perspective, (d,p) reactions are attractive since the
scattering problem can be viewed as an effective three-body
problem~\cite{ThompsonNunes}. One of the most challenging aspects of solving
the three-body problem for nuclear reactions is the repulsive Coulomb
interaction between the nucleus and the proton. While 
exact calculations of (d,p) reactions based on Faddeev equations in
the Alt-Grassberger-Sandhas (AGS)~\cite{ags} formulation can be carried
out~\cite{Deltuva:2009fp} for very light nuclei, this is not the case for 
heavier nuclei with
higher charges. The reason for this shortcoming is rooted in 
implementations of the Faddeev-AGS equations that rely on a screening
and renormalization procedure~\cite{Deltuva:2005wx,Deltuva:2005cc}, which
leads to increasing technical difficulties in computing (d,p) reactions with   
heavier nuclei~\cite{hites-proc}.

In Ref.~\cite{Mukhamedzhanov:2012qv}, a three-body theory for (d,p)
reactions is derived, where no screening of the Coulomb force is introduced. Therein, the Faddeev-AGS
equations are cast in the Coulomb-distorted partial-wave representation,
instead of the plane-wave basis. The interactions in the different two-body
subsystems, including the neutron- and proton-nucleus interactions, 
are assumed to be of separable form.

Separable forms for nucleon-nucleus interactions have been considered in the
past (e.g. \cite{Cattapan:1975tf,Cattapan:1975np}),
but are usually of a rank-1  Yamaguchi form and are intended to represent the
nuclear forces up to a few MeV.  
This is not sufficient for scattering of heavy nuclei up to tens of MeV. 
In addition, adjusting the parameters of Yamaguchi-type neutron-nucleus form factors
to obtain proton-nucleus form factors is not very practical when considering a larger variety
of nuclei. Therefore, a systematic scheme for deriving separable representations for
proton-nucleus optical potentials is needed.

In Ref.~\cite{Hlophe:2013xca} we derived a separable representation of phenomenological 
neutron-nucleus optical potentials, based on a generalization
of the Ernst, Shakin and Thaler (EST) scheme for non-hermitian interactions.
In Ref.~\cite{upadhyay:2014} we presented the first test calculations of
form factors in the momentum-space Coulomb basis, using the
neutron-nucleus interaction developed in~\cite{Hlophe:2013xca}.
In this work we generalize these studies for  proton-nucleus
interactions.

The derivations in the original EST work laid out in~\cite{Ernst:1973zzb} set up 
the scattering  problem in a complete plane-wave basis, whereas in
this work we need to use a  complete Coulomb basis. 
Consequently,  when working in momentum space,
we require a solution of the momentum space 
scattering equation in the Coulomb basis exists. 
We solve the momentum space Lippmann-Schwinger (LS) equation in the
Coulomb basis, following the method introduced in Ref.~\cite{Elster:1993dv} and successfully applied in 
proton-nucleus scattering calculations with microscopic optical
potentials in Ref.~\cite{Chinn:1991jb}. 
We note that a separable expansion for local potentials with Coulomb interactions
was first derived by Adhikari~\cite{Adhikari:1976zz}  and applied to 
proton-proton scattering. However, it has never been applied to proton 
scattering from heavier nuclei.
 
In Sec.~\ref{formal} we sketch the important steps needed to derive a separable
representation of a phenomenological global optical
potential in the momentum-space Coulomb basis  for proton-nucleus scattering. Our numerical calculations
of S-matrix elements for proton scattering from $^{48}$Ca and $^{208}$Pb at selected
laboratory kinetic energies are discussed in Sec.~\ref{results}, along
with the behavior of the form factors as a function of the external momentum. 
Finally, we summarize our work in Sec.~\ref{summary}. 


\section{Formal Considerations}
\label{formal}
\vspace{-3mm}

The scattering between a proton and a nucleus is governed by a potential
\begin{equation}
w = v^C + u^s ,
\label{eq:0a}
\end{equation}
where $v^C$ is the repulsive Coulomb potential and $u^s$ an arbitrary
short range potential. For the proton-nucleus system $u^s$ consists of 
an optical potential, which describes the nuclear interactions, and a
short-ranged Coulomb potential, traditionally parameterized as the potential of a charged sphere
with radius $R_0$ of which the point Coulomb force is subtracted~\cite{Varner:1991zz}.
Since the scattering problem governed by the point Coulomb force has 
an analytic solution, 
the scattering amplitude for elastic scattering between a
proton and a spin-zero nucleus is obtained as the sum of the Rutherford
amplitude $f^C(E_{p_0},\theta)$  and the Coulomb distorted nuclear amplitude given by
\begin{equation}
M^{CN} (E_{p_0},\theta) = f^{CN}(E_{p_0},\theta)+{\hat\sigma}\cdot
\mathbf{\hat{n}}\;  g^{CN}(E_{p_0},\theta), 
\end{equation}
with
\begin{eqnarray}
f^{CN}(E_{p_0},\theta)& =& -\pi\mu 
\sum_{l=0}^\infty e^{2i\sigma_l(E_{p_0})} P_l(\cos \theta) \times \cr
 \Big[ (l+1) \langle p_0 & |& \tau^{CN}_{l+}(E_{p_0})| p_0 \rangle + l
\langle p_0|\tau^{CN}_{l-}(E_{p_0})| p_0 \rangle\Big] ,  \\
 g^{CN}(E_{p_0},\theta)& =& -\pi\mu
\sum_{l=0}^\infty e^{2i\sigma_l(E_{p_0})} P^1_l(\cos \theta) \times \cr
 \Big[\langle p_0&|& \tau^{CN}_{l+}(E_{p_0})| p_0 \rangle - \langle p_0|\tau^{CN}_{l-}(E_{p_0})| p_0 \rangle\Big] .
\label{eq:0b}
\end{eqnarray}
Here $E_{p_0}=p_0^2/2\mu$ is the center-of-mass (c.m.) scattering energy which defines the
on-shell momentum $p_0$, and $\sigma_l=\;\arg\Gamma(1+l+i\eta)$ is the Coulomb phase shift.
The Sommerfeld parameter is given by $\eta=\alpha Z_1Z_2\mu/p_0$ with  $Z_1$ and $Z_2$ 
being the atomic numbers of the particles, and $\alpha$ the Coulomb coupling constant. 
The unit vector $\mathbf{\hat{n}}$ is normal to the scattering plane, and ${\hat\sigma}/2$ 
is the spin operator. The subscripts $'+'$ and $'-'$ correspond to a total angular momentum 
$j=l+1/2$ and $j=l-1/2$.  All calculations shown in this work refer to $j=l+1/2$.
Suppressing the total angular momentum indices for simplicity, 
the Coulomb distorted nuclear $t$-matrix element is given by 
$\langle p_0|\tau^{CN}_l(E_{p_0})| p_0 \rangle$, which is the solution of a LS type
equation, 
\begin{eqnarray}
\lefteqn{\langle p | \tau^{CN}_l(E_{p_0})| p_0 \rangle  = \langle p | u^s_l |
p_0 \rangle +} && \\
         & \int & p'^2 dp'  \langle p | u^s_l | p'\rangle \langle p'| g_c
(E_{p_0} +i\varepsilon)|p' \rangle \langle p' |\tau^{CN}_l(E_{p_0})| p_0
\rangle. \nonumber
\label{eq:0c}
\end{eqnarray} 
Here 
$g_c^{-1}(E_{p_0} +i\varepsilon)=E_{p_0}+ i\varepsilon -H_0 - v^C$ is the Coulomb 
Green's function, and $H_0$ the free Hamiltonian.
The Coulomb distorted nuclear $t$-matrix element $\langle p | \tau^{CN}_l(E_{p_0})| p_0 \rangle$
is related to the proton-nucleus $t$-matrix $\langle p|t_l(E_{p_0})|p_0\rangle$ by the 
familiar two-potential formula
\begin{eqnarray}
\lefteqn{ \langle p|t_l(E_{p_0})|p_0\rangle =} && \cr
&&\langle p|t^C_l(E_{p_0})|p_0\rangle+
 e^{2i\sigma_l(E_{p_0})}\langle p | \tau^{CN}_l(E_{p_0})| p_0 \rangle,
 \label{eq:0d}
\end{eqnarray}
where $\langle p|t^C_l(E_{p_0})|p_0\rangle$ is the point Coulomb $t$-matrix.

When the integral equation in Eq.~(\ref{eq:0c}) is solved in the basis of 
Coulomb eigenfunctions, 
$g_c$ acquires the form of a free Green's function and the difficulty of 
solving it is shifted to evaluating the potential matrix
elements in this basis.

For deriving a separable representation of the Coulomb distorted proton-nucleus $t$-matrix element,
we generalize the approach suggested by Ernst, Shakin, and Thaler (EST)~\cite{Ernst:1973zzb}, 
 to the charged particle case. The basic idea behind the EST
construction of a separable representation of a given potential is that 
the wave functions
calculated with this potential and the corresponding separable potential agree at given fixed
scattering energies $E_i$, the EST support points. The formal derivations of~\cite{Ernst:1973zzb} use the plane wave basis, which is standard for scattering involving
short-range potentials. However, the EST scheme does not depend on the basis and can equally
well be carried out in the basis of Coulomb scattering wave functions. 
In order to generalize the EST approach to charged-particle scattering, 
one needs to be able to obtain the scattering wave functions or half-shell
t-matrices from a given potential in the Coulomb basis, and then construct the corresponding
separable representation thereof.    

\subsection{The half-shell $t$-matrices in the Coulomb basis}
\label{subCoultm}
\vspace{-3mm}

In order to calculate the  half-shell $t$-matrix of Eq.~(\ref{eq:0c}), 
we evaluate the integral
equation in the Coulomb basis as suggested in~\cite{Elster:1993dv}, and note that 
in this case the Coulomb Green's function behaves like a free Green's function.  
Taking $|\Phi_{l,p}^{c}\rangle$ to represent the partial wave Coulomb eigenstate, 
the LS equation becomes 
\begin{eqnarray}
&& \langle \Phi_{l,p}^{c} |\tau^{CN}_l(E_{p_0})|\Phi_{l,p_0}^{c}\rangle =
\langle \Phi_{l,p}^{c} | u_l^s | \Phi_{l,p_0}^{c}\rangle  +  \cr
&& \int\limits_0^\infty \langle \Phi_{l,p}^{c} |u_l^s| \Phi_{l,p'}^{c} \rangle
\;\frac { p'^2 dp'}{E_{p_0} - E_{p'} +i\varepsilon} \langle
\Phi_{l,p'}^{c}|\tau^{CN}_l(E_{p_0})|\Phi_{l,p_0}^{c}\rangle  \cr
&& \equiv \langle p | \tau^{CN}_l(E_{p_0})| p_0 \rangle , 
\label{eq:1.1}
\end{eqnarray}
which defines the Coulomb distorted nuclear $t$-matrix of Eq.~(\ref{eq:0c}).
To determine the short-range potential matrix element, we follow Ref.~\cite{Elster:1993dv} and insert a complete set of position space eigenfunctions
\begin{eqnarray}
&&  \langle  \Phi_{l,p'}^{c} |u^s_{l}| \Phi_{l,p}^{c}\rangle= \frac{2}{\pi}\int\limits_0^\infty \langle\Phi_{l,p'}^{c}|r' \rangle
  \;r'^2dr\;\langle r'|u^s_{l}|r\rangle  \;r^2dr\;\langle r| \Phi_{l,p}^{c}\rangle \cr
  && = \frac{2}{\pi p' p} \int\limits_0^\infty r r' dr dr'  \;F_l(\eta',p'r')\;
  \langle r'| u^s_l |r \rangle  \;F_l(\eta,pr).
 \label{eq: 1.2}
 \end{eqnarray}
The partial wave Coulomb functions are given in coordinate space as
\begin{equation}
  \langle r| \Phi_{l,p}^{c}\rangle \equiv  \frac{e^{i\sigma_l(p)} F_l(\eta,pr)}{pr},
  \label{eq: 1.3}
\end{equation}
where $F_l(\eta,pr)$ are the standard Coulomb functions~\cite{AbramovitzStegun} 
and $\eta$($\eta'$) is the Sommerfeld parameter determined with momentum $p$($p'$).

 For our application we consider phenomenological optical potentials of Woods-Saxon
 form which are local in coordinate space. Thus the momentum space potential matrix elements simplify to
\begin{eqnarray}
\langle \Phi_{l,p'}^{c} |u^s_l| \Phi_{l,p}^{c} \rangle = 
\frac{2}{\pi p'p} \int\limits_0^\infty dr\;  F_l(\eta',p' r) u^s_l(r) F_l(\eta,pr).
\label{eq:1.6}
\end{eqnarray}
We compute these matrix elements for the short-range piece of the CH89 phenomenological
global optical potential~\cite{Varner:1991zz}, 
which consists of the nuclear and short range Coulomb potential. The
nuclear potential is parameterized using Woods-Saxon functions. For the short range Coulomb
interaction, the potential of a uniformly charged sphere is assumed, from which the point
Coulomb force is subtracted. The integral can be carried out with standard methods,
since $u_l^s(r)$ is short ranged and the coordinate space Coulomb wavefunctions are
well defined. The accuracy of this integral can be tested by replacing the Coulomb functions
with spherical Bessel functions and comparing the resulting matrix elements to the partial-wave 
decomposition
of the semi-analytic Fourier transform used in~\cite{Hlophe:2013xca}.
For the cases under study, and a maximum radius of 14~fm, 300 grid points
are sufficient to obtain matrix elements with a  precision of six significant digits.

\subsection{EST representation of the proton-nucleus $t$-matrix in the Coulomb basis}
\label{subCoulEST}
\vspace{-3mm}
 
 Extending the EST separable representation to the Coulomb basis involves replacing
 the neutron-nucleus half-shell $t$-matrix in Eqs.~(14)-(16) of Ref.~\cite{Hlophe:2013xca} 
 by the Coulomb distorted nuclear half-shell $t$-matrix. This leads to the  
 separable Coulomb distorted nuclear $t$-matrix
\begin{equation}
 \tau^{CN}_l(E_{p_0})=\sum_{i,j} u_l^s|f^c_{l,k_{E_i}}\rangle \;\tau^{c}_{ij}(E_{p_0})\;
 \langle {f^c}^*_{l,k_{E_j}}|u_l^s ,
 \label{eq:1.7}
\end{equation}
with $\tau^c_{ij}(E_{p_0})$  being constrained by
\begin{eqnarray}
&& \sum_i \langle {f^c}^*_{l,k_{E_n}}|u_l^s - u_l^s g_c(E_{p_0}) u_l^s| f^c_{l,k_{E_i}}\rangle \tau^{c}_{ij}(E) = \delta_{nj}  \\
&& \sum_j \tau^{CN}_{ij}(E_{p_0}) \; \langle {f^c}^*_{l,k_{E_j}}|u_l^s -u_l^s g_c (E_{p_0}) u_l^s|
f^{c}_{l,k_{E_k}}\rangle = \delta_{ik} \;. \nonumber
\label{eq:1.9}
\end{eqnarray}
Here $| f^c_{l,k_{E_i}}\rangle$  and $| {f^c}^*_{l,k_{E_i}}\rangle$ are the regular 
radial scattering wave functions corresponding to the short range potentials $u_l^s$  
and $(u_l^s)^*$ 
at energy $E_{i}$. The constraints of Eqs.~(\ref{eq:1.9}) ensure  that, at the EST support points, the exact and separable Coulomb
distorted nuclear half-shell $t$-matrices are identical. We want to point out that the generalization of the EST scheme to complex potentials~\cite{Hlophe:2013xca} is not 
affected by changing the basis from plane waves to Coulomb scattering states. 
The separable Coulomb distorted nuclear $t$-matrix elements are given by
 \begin{eqnarray}
&& \langle p'|\tau^{CN}_l(E_{p_0})|p\rangle \equiv \sum_{i,j}  h_{l,i}^c(p') \tau^c_{ij}(E_{p_0}) h_{l,j}^c(p) = \nonumber \\
&& = \sum_{i,j}\langle \Phi_{l,p'}^{c}| u_l^s| f^c_{l,k_{E_i}}\rangle \tau^c_{ij}(E_{p_0}) 
\langle {f^c}^*_{l,k_{E_j}} | u_l^s  | \Phi_{l,p}^{c}\rangle ,
\label{eq:1.10}
\end{eqnarray}
where the form factor 
\begin{eqnarray}
h_{l,i}^c(p) &\equiv & \langle \Phi_{l,p}^{c}| u_l^s| f^c_{l,k_{E_i}}\rangle \\
&=& \langle {f^c}^*_{l,k_{E_i}}|u_l^s| \Phi_{l,p}^{c}\rangle= 
\langle p|\tau^{CN}_{l}(E_i)| k_{E_i} \rangle \nonumber
\end{eqnarray}
is the short-ranged half-shell $t$-matrix satisfying Eq.~(\ref{eq:1.1}). 
For our analysis, and the comparison with coordinate-space calculations,
we consider the partial-wave $S$-matrix elements, which are obtained from the on-shell 
$t$-matrix elements by the relation \\
$  S_l(E_{p_0})= 1- 2\pi i \mu p_0\langle p_0|\tau^{CN}_l(E_{p_0})|p_0\rangle. $

Evaluating the separable Coulomb distorted proton-nucleus $t$-matrix involves integrals over 
the proton-nucleus form factor $h_{l,i}^c(p)$. If the short range Coulomb potential is omitted, 
the functional behavior of the proton-nucleus potential is similar to the one of the 
neutron-nucleus one,
and thus the numerical integration can be carried out as
discussed in~\cite{Hlophe:2013xca}. However, 
if it is included, the proton-nucleus form factor falls off more slowly as function of momentum. 
This implies that  larger maximum momenta and an increased number
of grid points are necessary to obtain a separable representation of the Coulomb distorted proton-nucleus 
$t$-matrix with the same accuracy as the separable representation of the neutron-nucleus
$t$-matrix.


\section{Results and Discussion}
\label{results}
\vspace{-3mm}

\begin{table}
\begin{tabular}{|r r r r| }
\hline\hline
$l$  & separable~~~ &   p-space~~~~  & r-space~~~~~\\[1ex]
\hline
 0 & -0.0512 0.3765  &-0.0518  0.3768 & -0.0523  0.3767 \\
 2 & 0.3805  0.0420  & 0.3809  0.0421 &  0.3808  0.0427\\
 6 &-0.0445  0.0170  &-0.0457  0.0118  & -0.0462  0.0111\\
10 & 0.9818 0.0248   & 0.9814  0.0253  &  0.9814  0.0253\\[1ex]
\hline
\end{tabular}
\caption{The partial wave S-matrix elements obtained from
the CH89~\cite{Varner:1991zz} phenomenological optical potential  for $j$= $l+1/2$ 
for selected 
angular momenta $l$ calculated for p$+^{48}$Ca elastic scattering at 
 $E_{lab}$~=~38~MeV.
}
\label{table-1}
\end{table}

For studying the quality of the separable  representation of $t$-matrices for proton-nucleus optical
potentials we consider 
p+$^{48}$Ca and p+$^{208}$Pb $S$-matrix elements in
the range of 0-50 MeV laboratory kinetic energy. We use the CH89 global optical 
potential~\cite{Varner:1991zz} and its rank-5 separable representation 
in all calculations. The same support points used 
for the neutron-nucleus separable representation (summarized in Table~I
of~\cite{Hlophe:2013xca}) 
provide a description of equal quality for the proton-nucleus $S$-matrix elements. This is
demonstrated for  p+$^{48}$Ca scattering at 38~MeV laboratory kinetic energy in 
Table~\ref{table-1}, which gives the $S$-matrix elements calculated with the separable representation of the Coulomb distorted
proton-nucleus $t$-matrix,  together with the corresponding direct 
calculations performed either in momentum  or coordinate space.

\begin{figure}
\begin{center}
\includegraphics[width=8.5cm]{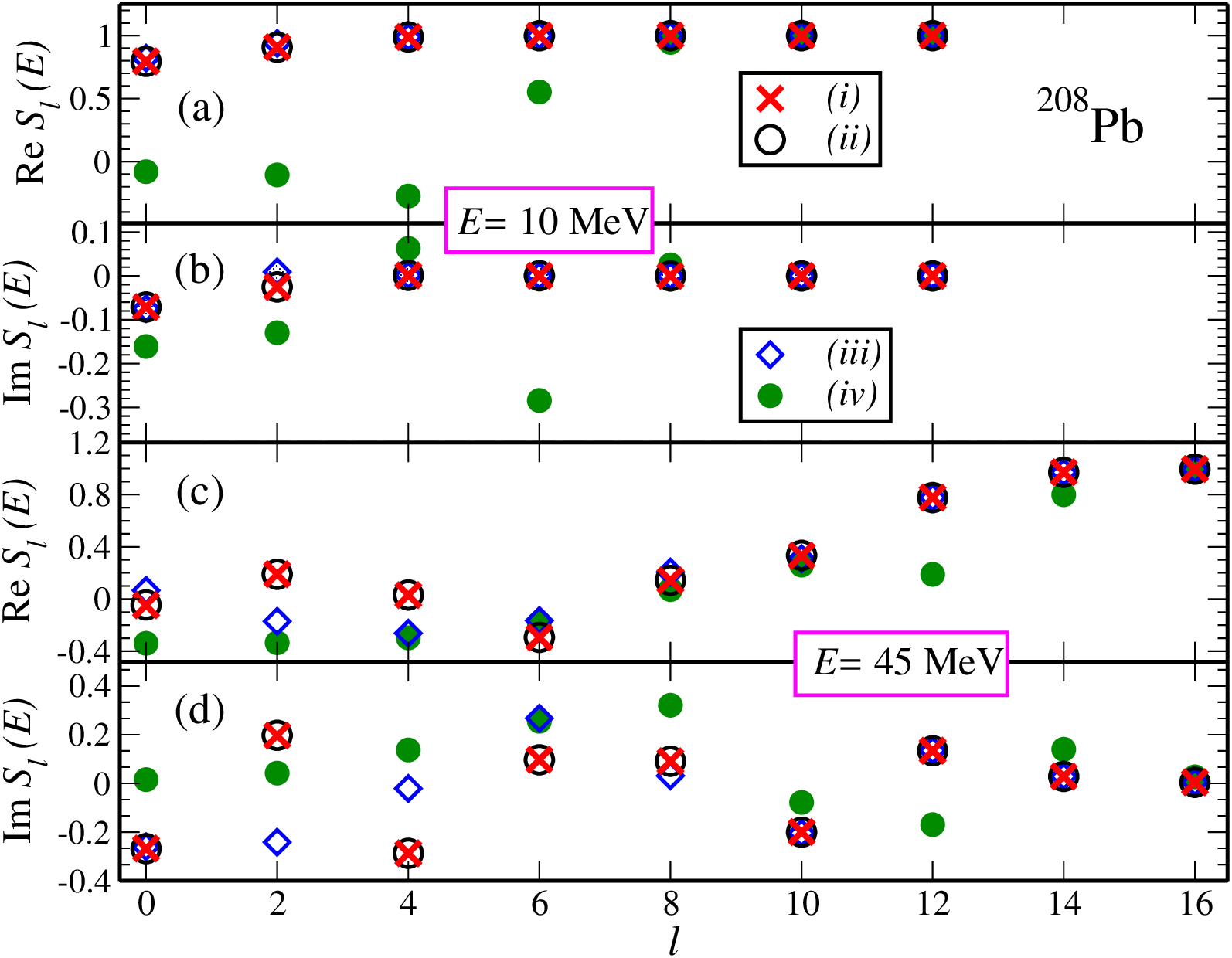}
\vspace{-1mm}
 \caption{(Color online)  The partial wave $S$-matrix for p+$^{208}$Pb elastic scattering 
 obtained from the CH89~\cite{Varner:1991zz} global optical potential as function of  
 angular momentum $j=$ $l+1/2$. Panels (a) and (b) show the real and
 imaginary parts of the $S$-matrix at $E_p=10$ MeV  and panels (c) and (d)
 provide the same information at $E_p= 45$ MeV: $(i)$  $S$-matrix elements
 calculated from the separable representation (crosses); $(ii)$  coordinate space calculation (open circles);
  $(iii)$  the calculation in which the short-range 
 Coulomb potential is omitted (open diamonds) and $(iv)$ 
 $S$-matrix elements for n+$^{208}$Pb elastic scattering (filled circles).}
\label{fig1}
\end{center}
\end{figure}

Similar results for the  p+$^{208}$Pb $S$-matrix elements are shown in Fig.~\ref{fig1}.
The top two panels (a) and (b) show the real and imaginary parts of the $S$-matrix elements 
at 10~MeV laboratory kinetic energy while the bottom two panels (c) and (d) show the real 
and imaginary parts of the $S-$matrix elements at 45~MeV. At 10~MeV the partial-wave series
converges much faster, thus we do not show matrix elements beyond $l=12$.
First, we note that the momentum space $S$-matrix elements calculated with the separable
representation (crosses) agree perfectly with the corresponding coordinate-space calculation (open circles). 

To illustrate the effects of the short-range Coulomb potential on the $S$-matrix elements, we show a calculation  in which
this term is omitted (open diamonds). As indicated in Fig.~\ref{fig1}, only the low $l$ partial waves are affected.
To demonstrate the overall size of all Coulomb effects for $^{208}$Pb, we also plot the 
corresponding n+$^{208}$Pb $S$-matrix elements at the same energies (filled circles). The differences between the crosses
and the filled circles demonstrate the importance of the correct inclusion of the Coulomb interaction.

\begin{figure}
\begin{center}
\includegraphics[width=8cm]{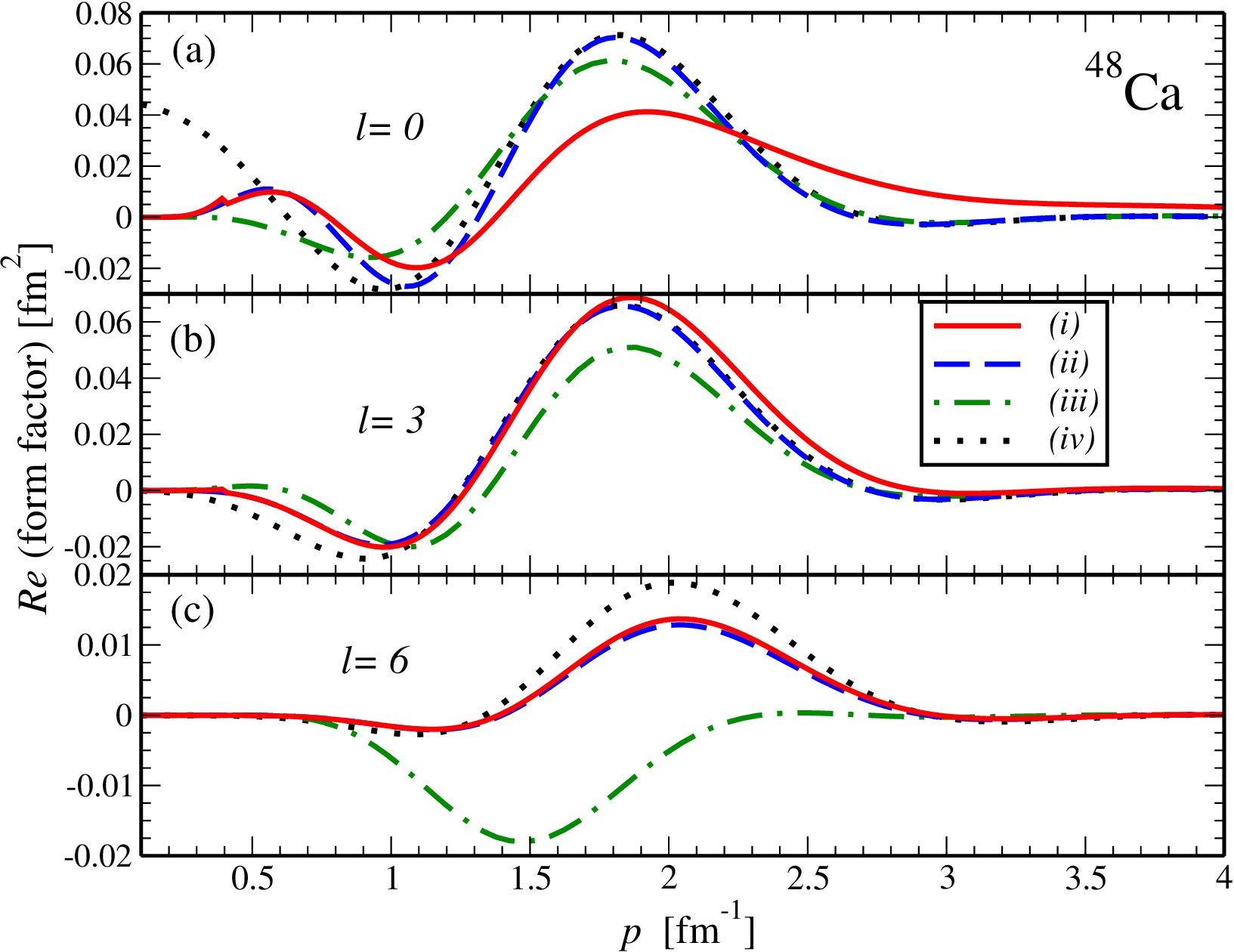}
\vspace{-1mm}
 \caption{ (Color online) The real parts of the partial wave proton-nucleus form factor
 for $^{48}$Ca as function of the momentum $p$ for selected angular momenta $l$:
 (a) $l=\;0$, (b) $l=\;3$, and (c) $l=\;6$. The form factors
 are calculated at $E_{c.m.}=\;36$ MeV and based on the CH89 global optical potential:
 full calculations ($i$) are compared to those omitting the short range Coulomb
($ii$), the neutron-nucleus form factor ($iii$)  and the Coulomb distorted
neutron-nucleus  form factor ($iv$). }
\label{fig2}
\end{center}
\end{figure}

\begin{figure}
\begin{center}
\includegraphics[width=8cm]{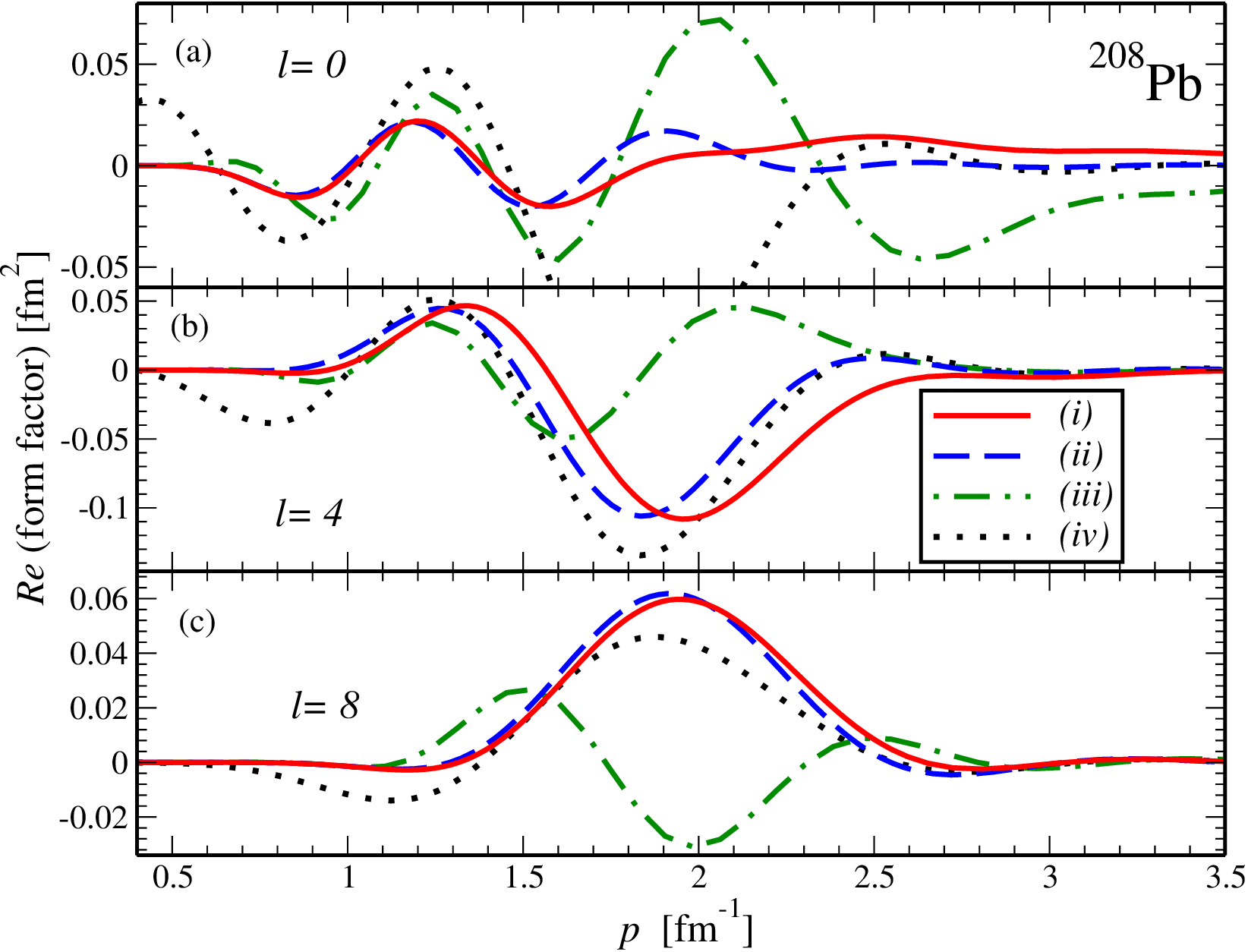}
\vspace{-1mm}
 \caption{(Color online). Same as Fig.~\ref{fig2} but for $^{208}$Pb. The form factors
 for $l=\;0$ (a) and $l=\;4$ are calculated at $E_{c.m.}=\;21$ MeV but for 
 $l=\;8$ (c) these are calculated at $E_{c.m.}=\;36$ MeV.
 }
\label{fig3}
\end{center}
\end{figure}

Next we examine the form factors of the separable representation in detail.
In Fig.~\ref{fig2} we compare p+$^{48}$Ca form factors for selected angular momenta  
calculated with the proton-nucleus potential and the short-range Coulomb potential ($i$)
to those calculated with the proton-nucleus potential alone ($ii$), as well as the n+$^{48}$Ca
($iii$). In addition we show the Coulomb distorted n+$^{48}$Ca form factor ($iv$) obtained
with the techniques introduced in~\cite{upadhyay:2014}  to
illustrate that a Coulomb distorted neutron-nucleus form factor differs from the corresponding Coulomb distorted
proton-nucleus form factor. 
First, we observe that, with the exception of $l=0$,
the form factors already vanish at 3.5 fm$^{-1}$.
For $l=0$, comparing the solid and dashed lines, we see that the short-range Coulomb 
potential significantly modifies the nuclear form factor.
The effects of the short-range Coulomb potential quickly decrease as $l$ increases. 

In Fig.~\ref{fig3}, we show a similar calculation for the $^{208}$Pb form factors. 
With the larger charge, the overall observations are maintained but magnified. 
For $l=0$, the short range Coulomb force creates a very slow
fall-off of the proton form factor, and only for $l=8$ is the
short-range Coulomb potential sufficiently weak to produce a negligible effect on the proton-nucleus form factor. 
Again we see that for  the angular momenta shown, the Coulomb distorted neutron-nucleus form factor does not resemble the
Coulomb distorted proton-nucleus form factor, emphasizing the need for a proper introduction of the Coulomb force in the EST scheme.

\section{Summary and Conclusions}
\label{summary}
\vspace{-3mm}
We have generalized the EST scheme~\cite{Ernst:1973zzb,Hlophe:2013xca} so that it can be applied to the scattering of charged particles with a repulsive Coulomb force.
To demonstrate the feasibility and accuracy of our method, we applied this Coulomb EST scheme to
elastic scattering of p+$^{48}$Ca and p+$^{208}$Pb. We found that the same EST
support points employed to obtain the neutron form factors can
be used for the separable representation of the  proton-nucleus potential. 
We showed that the momentum-space $S$-matrix elements calculated
with the separable representation of the Coulomb distorted proton-nucleus potential agree
very well with the corresponding coordinate-space calculation. Since
changing from a plane wave to a Coulomb basis preserves the time reversal 
invariance of the separable potential, the separable Coulomb
distorted proton-nucleus off-shell $t$-matrix also obeys reciprocity. 

We also studied the effects of the short-range Coulomb potential on the proton-nucleus
form factor. We found that, with the exception of the lowest partial waves ($l=$0, 1
for $^{48}$Ca and $l$= 0, 1, 2 for $^{208}$Pb), the form factors already vanish at 3.5 fm$^{-1}$.
For the lowest partial waves the short range Coulomb force creates a very slow fall-off 
for  the proton-nucleus form factor at high momenta.
The effects of the short-range Coulomb potential quickly decrease as $l$ increases and almost
vanish for $l$=6 ($^{48}$Ca) and $l=$8 ($^{208}$Pb). 

In addition, we demonstrated that 
the proton-nucleus form factor is very different from the Coulomb distorted neutron-nucleus form factor computed according to~\cite{upadhyay:2014}.  Thus, when applying those form factors in a
A(d,p)B Faddeev calculation, it will be mandatory to evaluate neutron and proton-nucleus
form factors separately.

\vfill


\acknowledgements
\vspace{-3mm}
This material is based on work  in part supported
by the U.~S.  Department of Energy, Office of Science of Nuclear Physics
under  program No. DE-SC0004084 and  DE-SC0004087 (TORUS Collaboration), under contracts DE-FG52-08NA28552  with
Michigan State University, DE-FG02-93ER40756 with Ohio University;  by  Lawrence Livermore
National Laboratory under Contract DE-AC52-07NA27344 and the U.T. Battelle LLC Contract DE-AC0500OR22725. 
F.M. Nunes  acknowledges support from the National Science Foundation
under grant PHY-0800026. This research used resources of 
the National Energy Research Scientific Computing Center, which is supported by the Office of
Science of the U.S. Department of Energy under Contract No. DE-AC02-05CH11231.

\vspace{-1mm}

\bibliography{coulomb}

\end{document}